\documentclass[twocolumn,longbibliography,aps,pra,preprintnumbers,citeautoscript]{revtex4-1}

\usepackage{graphicx}
\usepackage{bm}
\usepackage{color}
\usepackage{epstopdf}
\usepackage{amsmath}
\usepackage{amssymb}
\usepackage{epstopdf}
\usepackage{array}

\usepackage[urlcolor=blue,colorlinks=true,citecolor=blue,linkcolor=blue,pdfstartview={FitH},bookmarks=false]{hyperref}

\newcommand{\im}{\mathrm{i}}
\newcommand{\e}{\textrm{e}}
\newcommand{\ud}{\textrm{d}}

\begin{document}

\title{How to measure the Majorana polarization of a topological planar Josephson junction}

\author{Szczepan G\l{}odzik}
\email[e-mail:]{szglodzik@kft.umcs.lublin.pl}
\affiliation{Institute of Physics, M. Curie-Sk\l{}odowska University, 20-031 Lublin, Poland}

\author{Nicholas Sedlmayr}
\email[e-mail:]{sedlmayr@umcs.pl}
\affiliation{Institute of Physics, M. Curie-Sk\l{}odowska University, 20-031 Lublin, Poland}

\author{Tadeusz Doma\'nski}
\email[e-mail:]{doman@kft.umcs.lublin.pl}
\affiliation{Institute of Physics, M. Curie-Sk\l{}odowska University, 20-031 Lublin, Poland}

\date{\today}

\begin{abstract}
We analyze the topological superconductivity and investigate the spectroscopic properties manifested by the zero-energy modes, induced in a metallic strip embedded into a Josephson-type junction. Focusing on the Majorana polarization of such quasiparticles we propose feasible means for its empirical detection, using the spin selective Andreev reflection method. Our study reveals a gradual development of a transverse gradient of the Majorana polarization across the metallic strip upon increasing its width. We also inspect the spatial profile and polarization of the Majorana quasiparticles in the presence of a strong electrostatic impurity. We show that, depending on its position, such a defect can lead to a substantial localization of the Majorana mode.
\end{abstract}

\maketitle

\section{Introduction}
\label{sec.intro}

Topological materials, including those which are either insulators or superconductors, differ qualitatively from their ordinary counterparts due to the emergence of protected in-gap modes. Such quasiparticles, which develop at boundaries or internal defects, are topologically protected (thus being good candidates for stable qubits), and obey fractional statistics (which is appealing for quantum computations). Experimental efforts for the realization of these exotic quasiparticles have so far largely focused on one-dimensional structures, e.g.~semiconducting nanowires proximitized to superconductors~\cite{deng.yu.12,mourik.zuo.12,das.ronen.12,finck.vanharlingen.13,lutchyn.bakkers.18}, nanochains of magnetic atoms deposited on superconducting substrates~\cite{nadjperge.drozdov.14,
pawlak.kisiel.16,feldman.randeria.16,ruby.heinrich.17,jeon.xie.17,
kim.palaciomorales.18}, and lithographically fabricated nanostructures \cite{nichele.ofarrell.17}. Another direction in pursuit of topological superconductivity relies on two-dimensional systems, where the in-gap quasiparticles are chiral modes~\cite{rontynen.15,bjornson.15,li.16,rachel.mascot.17,Klinovaja_2018_chiral}. Such Majorana edge modes have indeed been observed in STM measurements, using nanoscopic islands of magnetic atoms deposited on superconducting substrates \cite{menard.17,He_etal.2017,wiesendanger_etal.18}. Further interesting perspectives are related with mixed-dimensionality systems, where the localized and delocalized Majorana quasiparticles coexist with one another \cite{Sedlmayr2016,PascalSimon_etal.18}. In particular, nanowires attached to larger structures~\cite{nichele.drachmann.17} could enable a controllable transfer of the Majorana modes between these constituents \cite{kobialka.ptok.domanski.19}, probing their Chern numbers \cite{Morr-2019}. 

\begin{figure}
\includegraphics[width=0.95\linewidth,keepaspectratio]{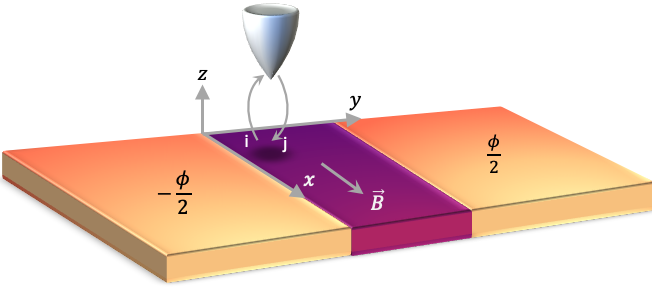}
\caption{A schematic view of a metallic strip (dark purple), embedded between superconducting regions (yellow) which differ in phase by $\phi$, probed by a polarized STM tip (light gray). A magnetic field $\vec B=B_0\hat x$ is applied to the whole structure.}
\label{schematics}
\end{figure}

Yet another promising platform for the realization of topological superconductivity hosting the Majorana modes has been suggested in Refs.~\cite{Pientka-2017,Hell-2017} using normal strips characterized by a strong spin-orbit coupling, embedded between two superconducting leads with differing phases (see Fig.~\ref{schematics}). Signatures of the zero-energy modes have already been reported for such heterostructures, consisting of aluminium on indium arsenide~\cite{Fornieri-2019} and an $HgTe$ quantum well coupled to a thin aluminium film~\cite{Ren-2019}. The major virtue of a Josephson-type geometry is its tunability to the topologically non-trivial regime, that can be  controlled experimentally by the phase difference. Another method for a controllable transition to the topological phase is possible by embedding two gate-tunable Josephson junctions in a phase-sensitive SQUID geometry, as reported for epitaxial $Al/InAs$ heterostructures \cite{Mayer-2019}. Experiments on these Josephson junction heterostructures~\cite{Fornieri-2019,Ren-2019,Mayer-2019} have triggered further intensive studies~\cite{SetiawanLevin-2019,Stern-2019_prb,Scharf-2019,Laeven-2019}. The proximitized strips are hoped, for instance, to enable a current-controlled braiding of the Majorana modes~\cite{Stern-2019_braiding}. It has also been suggested~\cite{Stern-2019_disorder} that weak disorder might promote localization of the Majorana quasiparticles.

Topological superconductivity of the phase controlled Josephson heterostructures has been observed spectroscopically in the narrow (albeit finite-width) normal strips placed between conventional superconductors. The group in Copenhagen used for this purpose a (1.6-5 $\mu$m long and 40-120 nm wide) $InAs$ region \cite{Fornieri-2019}, while in Harvard they explored (1-4 $\mu$m long and 400-600 nm wide) $HgTe$ quantum wells~\cite{Ren-2019}. In both cases dimensionality might have played an important role, affecting the character of the boundary modes. A convenient tool to deal with this issue is the {\em Majorana polarization}, introduced in Ref.~\cite{Sedlmayr2015b}, which is capable to inspect and identify topological nature of the zero-energy quasiparticles~
\cite{Sedlmayr2015b,Sedlmayr_etal_2015,Sedlmayr2016,maska.domanski.17}.

Formally the Majorana polarization is the local expectation value of the particle-hole operator for a given eigenstate ~\cite{Sedlmayr2015b}. It has been shown to be useful for calculating the topological phase diagram when alternative methods (specific either for one- 
or two-dimensional systems) cannot be adopted, and for recognizing whether the low energy states are the true Majorana quasiparticles~\cite{Sedlmayr2015b,Sedlmayr_etal_2015,Sedlmayr2016}.
In the present context, the quasi two-dimensionality of the proximitized strips is expected to induce transverse gradients in the Majorana polarization~\cite{Kaladzhyan2017a}. They would be observable in the phase of the Majorana polarization vector, and in its absolute value. Transverse gradients of the Majorana polarization, relative to the density, could be an indication of a delocalization process that ultimately might be detrimental to the zero-energy modes. Details concerning the Majorana polarization are presented in Sec.~\ref{polar_M}.

Furthermore, we prove (in Sec.~\ref{SESAR}) that the modulus of the Majorana polarization would be accessible experimentally using selective equal spin Andreev reflection spectroscopy~\cite{He-2014}. A similar technique has been already applied to detect the spin polarization of the Majorana quasiparticles of $Fe$ atom chains deposited on superconducting $Pb$~\cite{jeon.xie.17,maska.domanski.17} and to probe the zero-energy mode confined in a vortex core of a two-dimensional
$Bi_{2}Te_{3}/NbSe_{2}$ heterostructure \cite{Sun-2016}. The method proposed here could be an unambiguous probe of the Majorana nature of the zero-energy quasiparticles.


Finally, we address how robust the Majorana quasiparticles of the phase controlled Josephson junctions are against an electrostatic scattering potential placed in various regions of the proximitized strip. Our study reveals that, when this local defect is placed in an interior of the strip its influence on the Majorana modes is practically negligible, but when placed near a region of the existing Majorana quasiparticle we observe a tendency towards reducing the spatial extent of the zero-energy modes, analogous to what has been predicted by Haim and Stern in the disordered case~\cite{Stern-2019_disorder}. These phenomena are in stark contrast to the properties of Majorana quasiparticles of strictly 1-dimensional systems, where strong local defects usually produce additional pairs of the Majorana modes.

The paper is organized as follows. In Sec.~\ref{sec.model} we present the microscopic model and outline methodological details. Next, in Sec.~\ref{sec.transition}, we inspect the spatial profiles of the zero-energy modes and consider their Majorana polarization, focusing on their evolution upon varying the width of the proximitized strip. Sec.~\ref{SESAR} presents the selective Andreev spectroscopy and shows that it can probe the modulus of the {\em symmetrized} Majorana polarization. Sec.~\ref{sec.impurity} discusses the localization of the Majorana quasiparticle driven by a point-like electrostatic defect, and  Sec.~\ref{sec.sum} summarizes the main results.

\section{Microscopic model}
\label{sec.model}

For a schematic of the planar Josephson heterostructure, see Fig.~\ref{schematics}, here we employ the microscopic scenario discussed in Refs.~\onlinecite{Pientka-2017,Hell-2017,Scharf-2019}. The model tight binding Hamiltonian, 
\begin{equation}
\mathcal{H} = \mathcal{H}_{0} + \mathcal{H}_{Z} + \mathcal{H}_{S} \,,
\label{model_H}
\end{equation}
consists first of the free term
\begin{equation}
\mathcal{H}_{0} = \sum_{\substack{\langle i,j\rangle\\\sigma,\sigma'}} \left[
\im\lambda(\mathbf{d}_{ij} \times \vec{\bm{\sigma}}_{\sigma\sigma'})_z-t\delta_{\sigma\sigma'} \right] d_{i\sigma}^{\dagger} d_{j\sigma'} -\mu\sum_{i\sigma} d_{i\sigma}^{\dagger} d_{i\sigma} 
\end{equation}
describing itinerant electrons hopping all over the sample. $t$ is the hopping integral between the nearest neighbor atomic sites on a square lattice, $\lambda$ is the strength of the Rashba spin-orbit coupling, $\mathbf{d}_{ij}$ is the vector connecting nearest neighbors, and $\bm{\sigma}$ stands for the vector of the Pauli matrices. The second (Zeeman) term
\begin{equation}
\mathcal{H}_{Z} = 
 B_{0} \sum_{i}\sum_{\sigma\sigma'} d_{i\sigma}^{\dagger} \bm{\sigma}^x_{\sigma\sigma'}d_{i\sigma'}
\label{Zeeman_term}
\end{equation}
accounts for the influence of an external magnetic field $B_{0}$ which is parallel to the interface between the metallic and superconducting regions, as reported experimentally \cite{Fornieri-2019,Ren-2019}. The last part appearing in the model Hamiltonian \eqref{model_H} describes the  on-site pairing in the left ($S_{L}$) and right ($S_{R}$) superconducting regions, 
\begin{equation}
\mathcal{H}_{S} = 
\sum_{i} \left( \Delta_{i} d_{i\downarrow}^{\dagger} d_{i\uparrow}^{\dagger} + \mbox{\rm H.c.} \right)\,,
\end{equation}
where
\begin{equation}
\Delta_i = \left\{ \begin{array}{ll}
\Delta \e^{-\im\phi/2} &  \mbox{for } i \in S_{L}\,, \\
\Delta  \e^{ \im\phi/2} &  \mbox{for } i \in S_{R}\,,\textrm{ and}\\
0 &  \mbox{for }  i \in N\,.
\end{array} \right.
\end{equation}
Here the metallic strip region is denoted by $N$. The phase difference between the superconducting layers $S_{R}$ and $S_{L}$ is $\phi$ and $\Delta$ is real. 

We studied the finite-size version of this model, consisting of $N_x$ sites along the $x$-direction and $N_y$ sites along the $y$-direction. For specific computations we assumed $N_x=91$ and $N_y=30$, unless stated otherwise. The eigenstates and eigenenergies of the heterostructure were determined numerically, solving the Bogoliubov de-Gennes  equations with the canonical transformation %
\begin{equation} 
\left( \begin{array}{c} 
d_{i\uparrow}  \\ d_{i\downarrow}^{\dagger}
\end{array} \right) 
= \sum_{n} 
\left[ \begin{array}{cc} 
u_{i\uparrow}^{n} & v_{i\uparrow}^{n} \\
- ( v_{i\downarrow}^{n} )^{*} &  (u_{i\downarrow}^{n} )^{*}  
\end{array} \right] \;
\left( \begin{array}{c} 
\gamma_{n}  \\ \gamma_{n}^{\dagger}
\end{array} \right) \,,
\label{BdG_eqn}
\end{equation} 
where $\gamma_{n}^{(\dagger)}$ stand for the Bogoliubov quasiparticles which diagonalize the Hamiltonian: $H=\sum_{n}E_{n}\gamma_{n}^{\dagger}\gamma_{n} + \textrm{const}$.

\section{Topography of the Majorana modes}
\label{sec.transition}

Upon substituting the metallic strip between the superconducting reservoirs, their Cooper pairs leak into the normal region, inducing on-site electron pairing. This proximity effect is efficient nearby the bulk superconductors, up to distances smaller than the coherence length $\xi$. Here we consider metallic samples comprising a few $N_{w}$ atomic rows, whose spatial width $N_{w}a\leq\xi$, where $a$ is the inter-atomic distance. Under such a condition the proximity effect induces superconductivity across the entire metallic region. A fully self-consistent study of the pairing gap in all parts of this heterostructure has been discussed in Ref.~\cite{SetiawanLevin-2019}.

\begin{figure}[b!]
\includegraphics[width=0.99\linewidth,keepaspectratio]{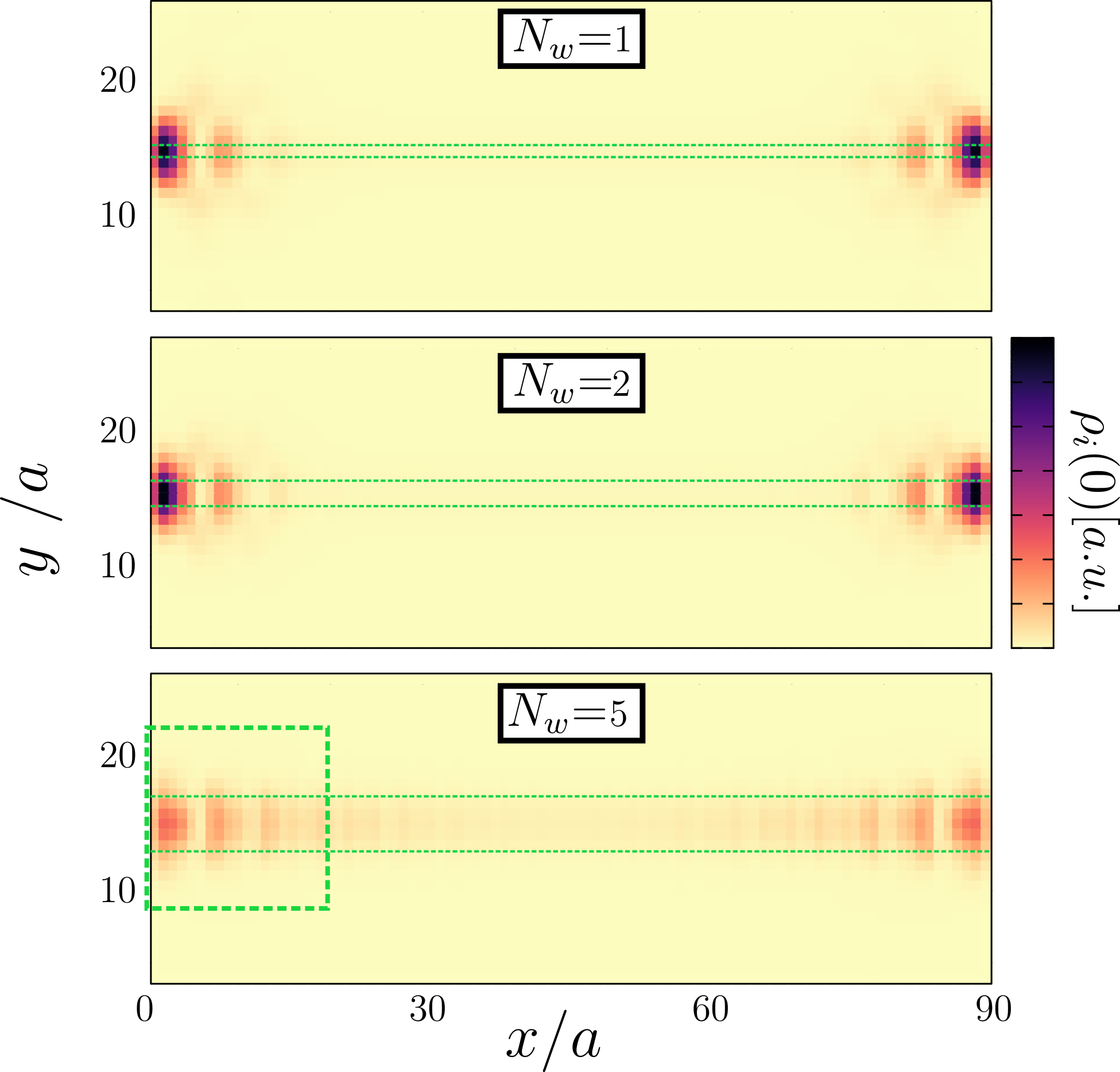}
\caption{The spatial profiles, $\rho_i(0)$, of the Majorana quasiparticles appearing in a metallic strip consisting of 1, 2 and 5 rows of atomic chains, as indicated. We have used the model parameters $\Delta=0.25t$, $\phi=\pi$, $\lambda=0.5t$, $B_{0}=0.1t$, $\mu=-3.75t$. The Majorana polarization \eqref{polarization_per_spin} for the area in the green square is shown in Fig.~\ref{polarization_MQD}.} 
\label{homogeneous}
\end{figure}

The topological superconducting phase originates from the triplet pairing, which can be achieved by combining the on-site pairing with the spin-orbit Rashba interaction and Zeeman splitting \cite{lutchyn.bakkers.18}.
It has been demonstrated \cite{Pientka-2017,Hell-2017,SetiawanLevin-2019,Stern-2019_prb,Scharf-2019,Laeven-2019} that a transition from the topologically trivial to the nontrivial superconducting state is sensitive to the Josephson phase $\phi$. Characteristic features of the emerging Majorana quasiparticles can, however, additionally depend on the width $N_{w}a$ of the metallic strip. In what follows we analyze such qualitative changes observable in the spectral function and the Majorana polarization vector. We also propose a method for empirical detection of the Majorana polarization (Sec.~\ref{SESAR}).

\subsection{Zero-energy spectral function}

Focusing on the optimal condition for the topological superconducting state $\phi=\pi$, 
we have checked that the on-site pairing $\left< d_{i\downarrow} d_{i\uparrow}\right>$ 
spreads nearly uniformly onto the metallic strip, both along and across it. Furthermore, we also noticed some feedback of the metallic sector onto superconducting regions manifested by partial reduction of the local pairing, sometimes referred to as the inverse superconducting proximity effect. 

In presence of the spin-orbit interaction and the Zeeman field, the proximitized strip develops inter-site pairing of identical spin electrons, i.e.~triplet pairing. For sufficiently strong magnetic fields $B_{0}$, such a triplet superconducting phase becomes topologically nontrivial, leading to the emergence of the zero-energy quasiparticles~\cite{Pientka-2017,Hell-2017}. Their signatures can be observed  in the local density of states
\begin{equation}
\rho_{i}(\omega)=\sum_{n,\sigma} \left[ |u_{i\sigma}^{n}|^{2} \delta( \omega - E_{n})
+ |v_{i\sigma}^{n}|^{2} \delta( \omega + E_{n}) \right] \,.
\label{spectral}
\end{equation} 
As we consider a finite size system we have broadened the delta peaks to Lorentzian functions of width $0.02\Delta$.

\begin{figure}[b!]
\includegraphics[width=0.99\linewidth,keepaspectratio]{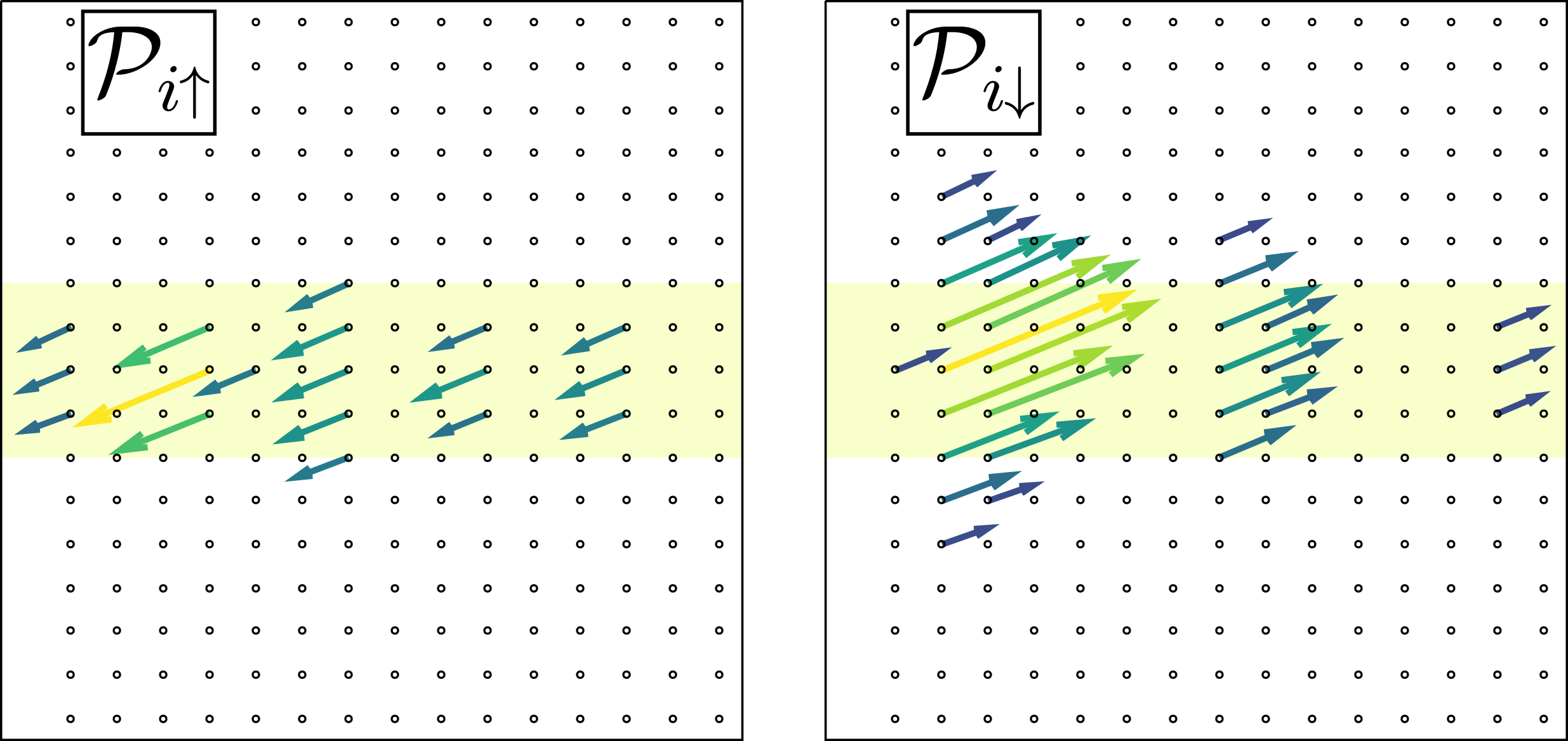}
\caption{Components of the Majorana polarization $\mathcal{P}_{i\sigma}$  obtained for the region highlighted by the dashed frame in Fig.~\ref{homogeneous}. The magnitude of the arrows shows $|\mathcal{P}_{i\sigma}|$ and their direction shows $\textrm{Arg}\,\mathcal{P}_{i\sigma}$. We note that the phase of the Majorana polarization is only well defined up to a global shift. The shaded region is the metallic strip.} 
\label{polarization_MQD}
\end{figure}

\begin{figure*}
\includegraphics[width=0.99\linewidth,keepaspectratio]{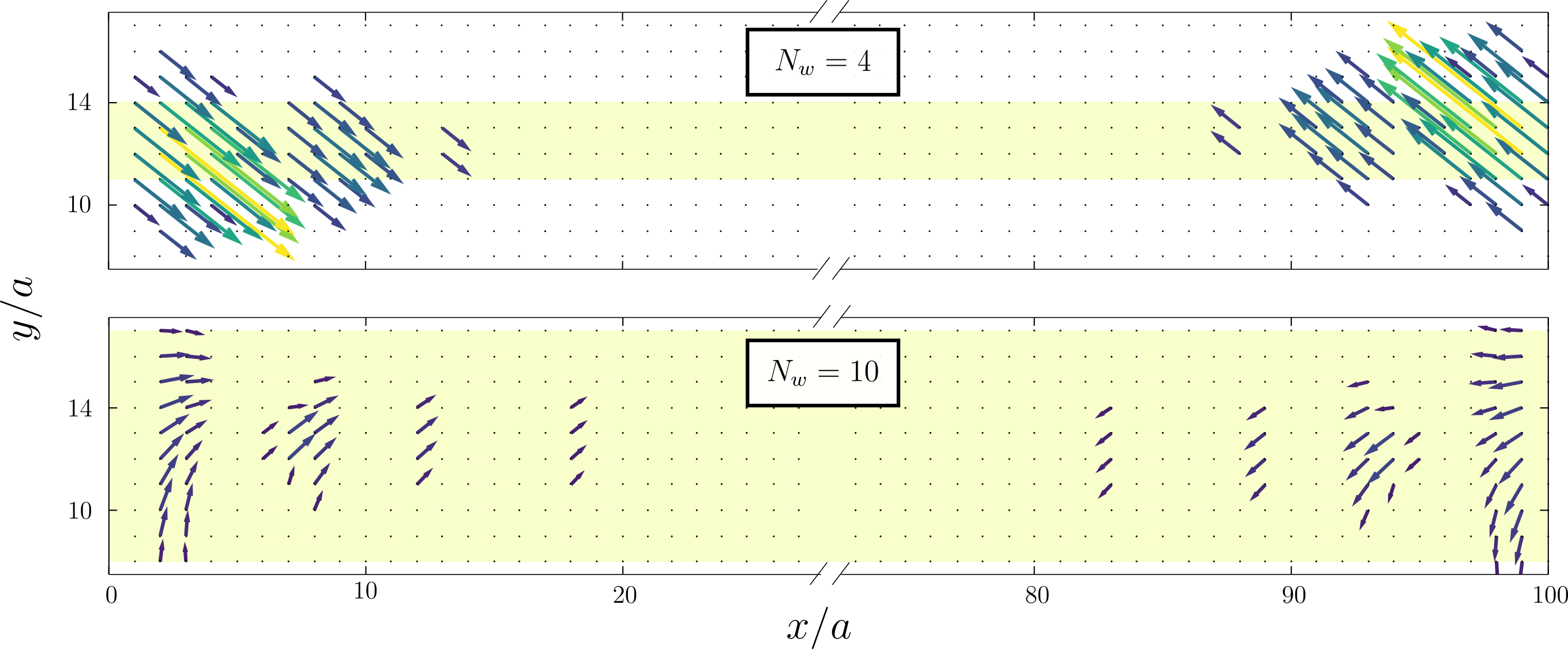}
\caption{The Majorana polarization $\mathcal{P}_{i}$ obtained for the heterostructure comprising $N_{w}=4$ (top) and $N_{w}=10$ (bottom) atomic rows in the metallic strip, marked by the shaded region. Numerical results are obtained for the same model parameters as in Fig. \ref{homogeneous} but using $N_{x}=100$, $N{y}=20$. The magnitude of the arrows shows $|\mathcal{P}_{i\sigma}|$ and their direction shows $\textrm{Arg}\,\mathcal{P}_{i\sigma}$. We note that the phase is only well defined up to a global shift.} 
\label{polarization_wide}
\end{figure*}

Fig.~\ref{homogeneous} displays the spatial profiles of the local density of states at zero energy $\rho_i(0)$, 
obtained for very narrow metallic strips. We note that the Majorana quasiparticles of the narrow metallic strip are well localized at its ends. Their overall topography is practically identical with all features of one-dimensional systems, including the  characteristic oscillations along the metallic strip \cite{Klinovaja_2016_tomography}. It comes as perhaps some surprise that this narrow width of metallic region is neither essential for the development of the topological superconducting phase, nor important for the spatial profile of the Majorana modes. Even in the extreme case $N_{w}=0$, i.e.~without any metallic piece between the phase-differing superconductors, such modes are still present. On the other hand, when the width $N_{w}$ increases we see a gradual smearing of the zero-energy quasiparticles. This is a consequence of the reduced proximity induced gap in wider strips which naturally reduces the localization of any mid-gap states.

\subsection{Majorana polarization}
\label{polar_M}

Majorana modes are quasiparticles with energy $E_{n}=0$ (we denote such doubly-degenerate eigenstate by $n\equiv n_{0}$) and which are eigenstates of the particle-hole transformation operator. We analyze here another valuable source of information about these modes encoded in the Majorana 
polarization~\cite{Sticlet_Bena_Simon-2015,Sedlmayr2015,Sedlmayr_etal_2015,Sedlmayr2015b}. This quantity is particularly useful for characterizing the Majorana modes of quasi two-dimensional topological superconductors~\cite{Sedlmayr2015b,Sedlmayr2016,Kaladzhyan2017a} where its phase develops both longitudinal and transverse variation. As we shall see, its texture brings an important message about the delocalized Majorana quasiparticles.

Formally, the particle-hole overlap can be defined for any eigenstate $|\psi_n\rangle$~\cite{Sedlmayr2015b,Sedlmayr2016}
\begin{equation}
    \mathcal{P}_{in}=\langle\psi_n|\mathcal{C}\hat{r}_i|\psi_n\rangle=\sum_\sigma\bm{\sigma}^z_{\sigma\sigma}2u_{i\sigma}^{n}v_{i\sigma}^{n}\,,
\end{equation}
where $\hat{r}_i$ is projection onto site $i$ and $\mathcal{C}$ is the particle-hole operator. More generally one may wish to consider the particle-hole overlap $u_{i\sigma_{1}}^{n}v_{j\sigma_{2}}^{m}$. In particular, the equal-spin pairing ($\sigma_{1}=\sigma_{2}$) induced between the neighboring sites $i$ and $j$ for the zero-energy quasiparticles ($E_{n}=0=E_{m}$) would be of our interest here. For convenience we introduce the local Majorana polarization
\begin{equation}
{\cal{P}}_{i} = {\cal{P}}_{i\uparrow} - {\cal{P}}_{i\downarrow} 
\label{Majorana_polarization}
\end{equation} 
where 
\begin{equation}
{\cal{P}}_{i\sigma} = 2 u_{i\sigma}^{n_0} v_{i\sigma}^{n_0} \,.
\label{polarization_per_spin}
\end{equation}
This complex quantity (\ref{Majorana_polarization}) allows one to probe the Majorana eigenstates $E_{n_{0}}=0$.  

Let us start by checking the contributions ${\cal{P}}_{i\sigma}$ from each spin $\sigma$ to the Majorana polarization in the narrow metallic strips, when the zero-energy quasiparticles 
are well localized near its ends. To be specific, we focus on the region marked by the dashed lines in the bottom panel in Fig.~\ref{homogeneous}. Both constituents ${\cal{P}}_{i\sigma}$ 
are depicted in Fig.~\ref{polarization_MQD} on the lattice sites of the marked metallic region. We clearly note that the directions of the arrows depicting ${\cal{P}}_{i\uparrow}$ are opposite to ${\cal{P}}_{i\downarrow}$, which is typical for strictly one-dimensional topological superconductors (see Fig.~2 in Ref.~\onlinecite{maska.domanski.17}) and for higher dimensions as well~\cite{Sedlmayr2015b,Sedlmayr2016,Kaladzhyan2017a}. The magnitudes of the two components, shown by the length of the arrows, are however very different. It is important to emphasize, that for the realization of a true MBS the local phase of $\mathcal{P}_{i\uparrow}-\mathcal{P}_{i\downarrow}$ must be constant. Such a constraint $\mathcal{P}_{i\uparrow}=\e^{\im\varphi}|\mathcal{P}_{i\uparrow}|=-\e^{\im\varphi}|\mathcal{P}_{i\downarrow}|$ seems to be satisfied in our case only for narrow metallic strips. 

There are in fact two conditions on $\mathcal{P}_i$ which are required for Majorana modes. The first, which we have seen here, is that its phase must be constant. The phase of $\mathcal{P}_i$ does not appear to be measurable in any simple way. What is measurable, as we will show in Sec.~\ref{SESAR}, is its modulus $|\mathcal{P}_i|$. For a Majorana mode we require $|\mathcal{P}_i|=\rho^0_i$, and this gives a measurable determination of Majorana modes. Upon increasing the width of the metallic strip the Majorana polarization gradually develops varying orientations, both along and across the sample. Emergence of the transverse gradient is very sensitive to the width $N_{w}$, as illustrated in Fig.~\ref{polarization_wide}.
We thus observe that the Majorana polarization vector has more subtle structure in comparison to the spectral function 
$\rho_{i}(0)$. In the next section we shall discuss empirical means to probe this quantity.

\section{Polarized Andreev spectroscopy}
\label{SESAR}

Here we discuss an empirical method based on spin polarized Andreev reflection spectroscopy \cite{He-2014}, which could probe the absolute value of the Majorana polarization. Let us consider a scanning tunnelling microscope (STM) tip  brought in contact with the site $j$ of our heterostructure. The influence of this external reservoir of itinerant electrons can be incorporated by augmenting the model Hamiltonian 
Eq.~\eqref{model_H} with the local term
\begin{equation}
H_{j} = \underbrace{\sum_{{\bf k},\sigma} (\varepsilon_{\bf k}-\mu_{tip}) 
c_{{\bf k} \sigma}^{\dagger} c_{{\bf k} \sigma}}_{\textrm{STM tip}} 
+ \underbrace{\sum_{{\bf k},\sigma} \left( t_{{\bf k},\sigma} 
c_{{\bf k}\sigma}^{\dagger} d_{j\sigma} + H.c.\right)}_{\rm hybridization} \,,
\label{tip_hamil}
\end{equation}
where the chemical potential of the tip, $\mu_{tip}=\mu+eV$, can be varied by a bias potential $V$. $\varepsilon_{\bf k}$ is the dispersion of the tip electrons and $t_{{\bf k},\sigma}$ denotes the tunneling amplitude from the tip to the heterostructure and vice-versa. The quasiparticle spectrum at site $j$ can be indirectly inferred from measurements of the charge transport induced by the voltage $V$ applied between the STM tip and the sample. In the subgap regime, i.e.~for $e|V| \leq \Delta$, such  current is contributed solely by the Andreev scattering processes. This mechanism relies on the conversion of electrons arriving from the STM tip into the Cooper pairs of the superconducting heterostructure, reflecting holes back into the STM tip. 

Since we are interested to probe the topological superconducting phase related to the intersite pairing of identical spin electrons, similar to the Kitaev scenario \cite{kitaev.01}, let us assume a complete polarization of the tip electrons. Under such circumstances only one spin component participates in the charge transport. On a microscopic level we can thus imagine  an electron of spin $\sigma$ arriving from the polarized STM tip at site $j$, where it forms the triplet pair with another electron of the same spin from the neighbouring site $i$, reflecting a hole back into the tip. 

The charge  flow from the STM tip can be defined as $I^{\sigma}(V) = -e \left< \frac{d}{dt} 
\sum_{\bf k} c_{{\bf k} \sigma}^{\dagger} c_{{\bf k} \sigma} \right>$. Using the Heisenberg equation 
of motion we can recast it in terms of the lesser Green's function $I^{\sigma}(V)=\frac{2e}{\hbar} 
\sum_{\bf k} \mbox{\rm Re} \left\{ t_{\bf k} \langle \langle d_{j\sigma} ; c_{{\bf k},\sigma}^{\dagger} 
\rangle\rangle^{<} \right\}$ whose determination is feasible within the Keldysh approach. Let us remark, 
however, that due to the intersite (triplet) pairing the operator $d_{j\sigma}$ would be coupled
to the neighboring site operator $d_{i\sigma}$. For this reason such current contributed by the Andreev
scattering through the sites $j$ and $i$ (as illustrated in Fig.~\ref{schematics}) will be explicitely 
denoted by subindices $I^{\sigma}(V) \equiv I_{ij}^{\sigma}(V)$. 

Following the steps, outlined previously by one of us in Ref.~\cite{maska.domanski.17}, the spin-polarized 
Andreev current can be expressed by the popular Landauer-type formula
\begin{equation}
I_{ij}^{\sigma}(V) = \frac{e}{h} \int \ud\omega \; T_{ij}^{\sigma}(\omega) \; 
\left[ f(\omega \!-\! eV) \!-\! f(\omega\!+\!eV) \right] \,,
\label{Andreev-current}
\end{equation}
where $f(\omega)=\left[1+\mbox{\rm exp}\left(\omega/k_{B}T \right)\right]^{-1}$
is the Fermi-Dirac distribution function. The main quantities of our interest would be
the spatially-dependent transmission probabilities, characterizing conversion 
of electrons into holes on the neighbouring sites~\cite{maska.domanski.17}
\begin{equation} 
T_{ij}^{\sigma}(\omega) = \Gamma_{N}^{2} \;  \left| 
{\cal{F}}_{ij}^{\sigma}(\omega) \right|^{2} \,,
\label{unconventional}
\end{equation} 
where ${\cal{F}}_{ij}^{\sigma} (\omega) = \langle\langle \hat{d}_{i\sigma};\, \hat{d}_{{j}\sigma} 
\rangle\rangle_{\omega}$ is the Fourier transform of the off-diagonal (in Nambu representation) 
retarded Green's function. For practical reasons (since we focus on a narrow transport window being 
a fraction of meV around the chemical potential $\mu$) we have introduced a constant coupling strength, 
$\Gamma_{N}\equiv 2\pi\sum_{\bf k} |\gamma_{{\bf k}}|^{2} \delta (\omega - \varepsilon_{\bf k})$. 
Formally this is equivalent to the wide band limit approximation. 

In the absence of the STM tip the Green's function  ${\cal{F}}_{ij}^{\sigma} (\omega)$ 
can be found explicitly from the Bogoliubov de Gennes diagonalization (\ref{BdG_eqn}). 
In particular, for $\sigma\!=\!\uparrow$ the transformation $d_{i\uparrow}=\sum_{n}
\left[ u_{i\uparrow}^{n} \gamma_{n}+ v_{i\uparrow}^{n} \gamma^{\dagger}_{n} \right]$  implies
\begin{eqnarray} 
\lim_{t_{\bf k,\sigma}\!=\!0}
\langle\langle d_{i\uparrow} ; d_{j\uparrow} \rangle\rangle_{\omega} =
\sum_{n} \left[ \frac{u_{i\uparrow}^{n}   v_{j\uparrow}^{n}  }{\omega-E_{n}}
+ \frac{v_{i\uparrow}^{n}u_{j\uparrow}^{n} }{\omega+E_{n}} \right] 
\end{eqnarray} 
and similar expression (with minus sign) holds for $\sigma=\downarrow$. 
The effect of the hybridization term introduced in Eq.~(\ref{tip_hamil}) can be modelled
in the wide-band limit approximation by the substitution $\omega \rightarrow \omega 
+ i\frac{\Gamma_{N}}{2}$ (see Appendix A in Ref.~\cite{maska.gorczyca.17} for a detailed derivation).
Physically it means, that the STM tip gives rise to a broadening of the subgap quasiparticles,
i.e.~to a finite life-time.
The Green's function ${\cal{F}}_{ij}^{\uparrow} (\omega)$ is thus given by
\begin{equation} 
{\cal{F}}_{ij}^{\uparrow} (\omega)
=  \sum_{n} \left[ 
\frac{u_{i\uparrow}^{n}   v_{j\uparrow}^{n}  }{\omega-E_{n}+ i\frac{\Gamma_{N}}{2}}
+ \frac{u_{j\uparrow}^{n} v_{i\uparrow}^{n}}{\omega+E_{n}+ i\frac{\Gamma_{N}}{2}}
\right]\,,
\end{equation} 
where the complex coefficients $u_{j\uparrow}^{n}$ and $v_{j\uparrow}^{n}$ have 
to be obtained numerically from the diagonalization procedure. 

\begin{figure}[t!]
\includegraphics[width=0.95\linewidth,keepaspectratio]{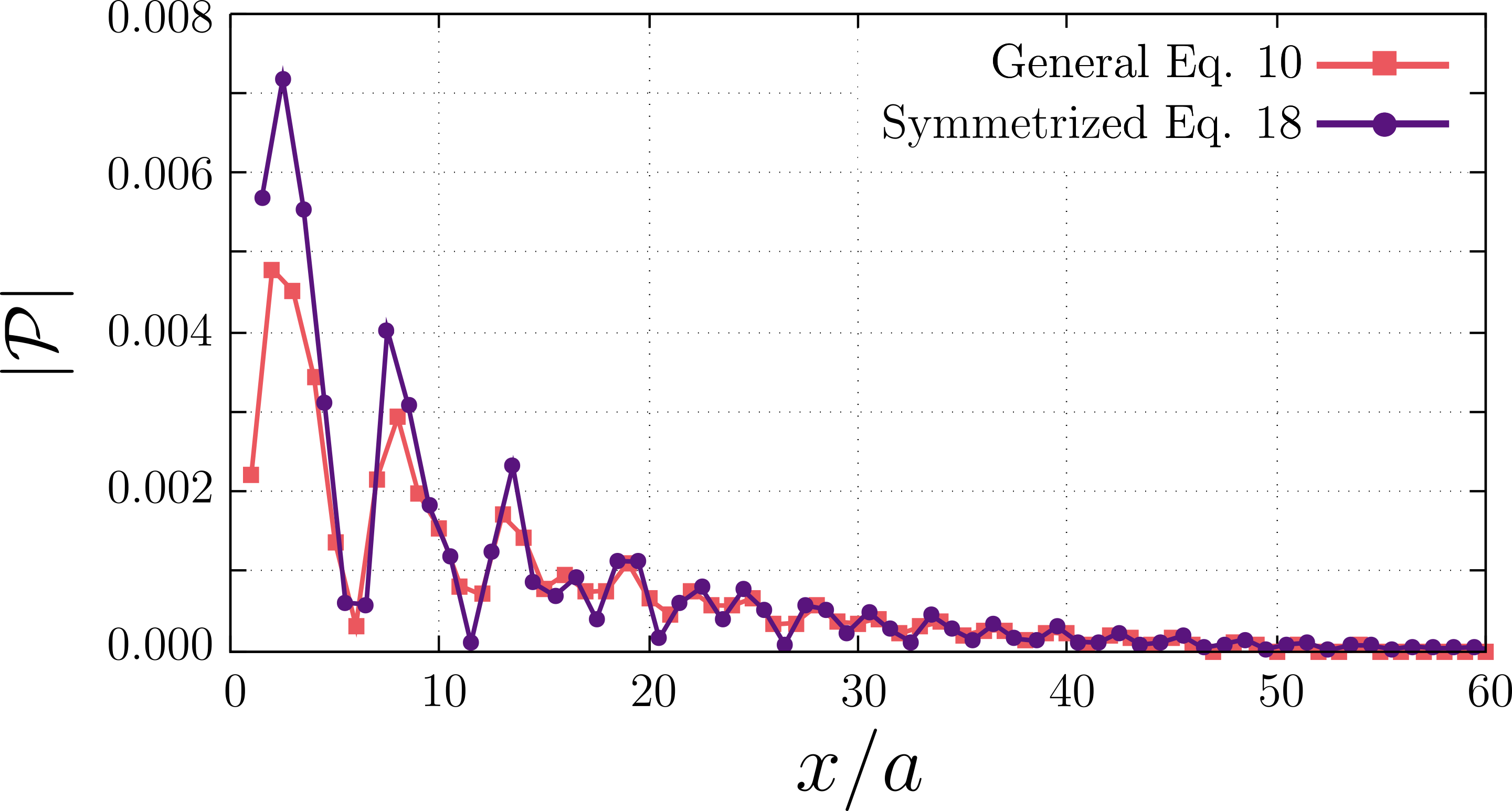}
\caption{Comparison of the Majorana polarization (\ref{Majorana_polarization}) with its symmetrized varsion defined in Eq.~(\ref{averaged_polarization_per_spin}) computed for the central row of the metallic strip using $N_{w}=5$ and the same model parameters as presented in Fig.~\ref{fig_rat}.} 
\label{profiles}
\end{figure}

Focusing on the zero-energy limit $\omega \rightarrow 0$, dominated by the Andreev scatterings 
via the Majorana quasiparticle (denoted here by $E_{n_{0}}=0$), we obtain the transmittance 
(\ref{unconventional}) simplified to
\begin{equation} 
T_{ij}^{\sigma}(\omega = 0 ) \simeq 
\left| 2 u_{i\sigma}^{n_0} v_{j\sigma}^{n_0} 
+ 2 u_{j\sigma}^{n_0} v_{i\sigma}^{n_0} \right|^{2}\,.
\label{off-diag}
\end{equation} 
Eq.~\eqref{off-diag} substituted into the spin-resolved Andreev current formula \eqref{Andreev-current} 
yields, at low temperatures, the following zero-bias differential conductance
\begin{equation} 
\lim_{V\rightarrow 0} \frac{d I_{ij}^{\sigma}(V)}{dV} \simeq \frac{4e^{2}}{h}
\left| u_{i\sigma}^{n_0} v_{j\sigma}^{n_0} 
+ u_{j\sigma}^{n_0} v_{i\sigma}^{n_0}\right|^{2} \,.
\label{zero-bias}
\end{equation} 
This  result demonstrates, that the polarized Andreev spectroscopy 
could probe the spin-dependent contribution \eqref{polarization_per_spin} to the Majorana polarization. 
Strictly speaking, however, such tunneling processes occur on the links (involving the neighbouring 
sites  $j$ and $i$) rather than on individual local sites. For this reason the differential 
conductance \eqref{zero-bias} would measure the {\em symmetrized  Majorana polarization}
\begin{equation}
{\cal{P}}_{<ij>,\sigma} =  u_{i\sigma}^{n_{0}} v_{j\sigma}^{n_{0}} + 
u_{j\sigma}^{n_{0}} v_{i\sigma}^{n_{0}} 
\label{averaged_polarization_per_spin}
\end{equation} 
over the neighboring sites $i$ and $j$, instead of the strictly local definition \eqref{polarization_per_spin}. Since the diagonalization coefficients are slowly varying in space, $u_{j\sigma}^{n}\approx u_{i\sigma}^{n}$, $v_{j\sigma}^{n} \approx v_{i\sigma}^{n}$, the symmetrized ${\cal{P}}_{<ij>,\sigma}$ and local ${\cal{P}}_{i,\sigma}$ Majorana polarizations are fairly similar (see Fig.~\ref{profiles}). Some discrepancy appears at the most peripheral sites, probably due to the boundary conditions.

Let us finally recall that the complex vector $u_{i\uparrow}^{n_0} v_{i\uparrow}^{n_0}$ is typically perfectly opposite to $u_{i\downarrow}^{n_0} v_{i\downarrow}^{n_0}$, see  Fig.~\ref{polarization_MQD}. Such anti-alignment implies, that a modulus of the Majorana polarization can be expressed as \eqref{Majorana_polarization} $\left| {\cal{P}}_{i} \right| = \left| {\cal{P}}_{i\uparrow} - {\cal{P}}_{i\downarrow} \right| = \left| {\cal{P}}_{i\uparrow} \right|\; + \; \left| {\cal{P}}_{i\downarrow} \right| $, and the same holds for the symmetrized Majorana polarization.
By measuring the zero-bias conductance $\frac{d I_{ij}^{\sigma}(0)}{dV}$ of the spin-selective Andreev current flowing through the neighboring sites $i$ and $j$ one can thus evaluate the absolute value of the symmetrized Majorana polarization
\begin{equation}
\left| {\cal{P}}_{<ij>} \right| = 
\sqrt{\frac{h}{4e^{2}}} \; \sqrt{\frac{d}{dV}  \left[ I_{ij}^{\uparrow}(V)+
I_{ij}^{\downarrow}(V) \right]_{V=0}} \,.
\label{polarization_experiment}
\end{equation} 
As far as the spatial variation of the phase is concerned, its determination would be evidently more cumbersome. This problem is beyond the scope of the present study.

\begin{figure}[t!]
 \includegraphics[width=0.95\linewidth,keepaspectratio]{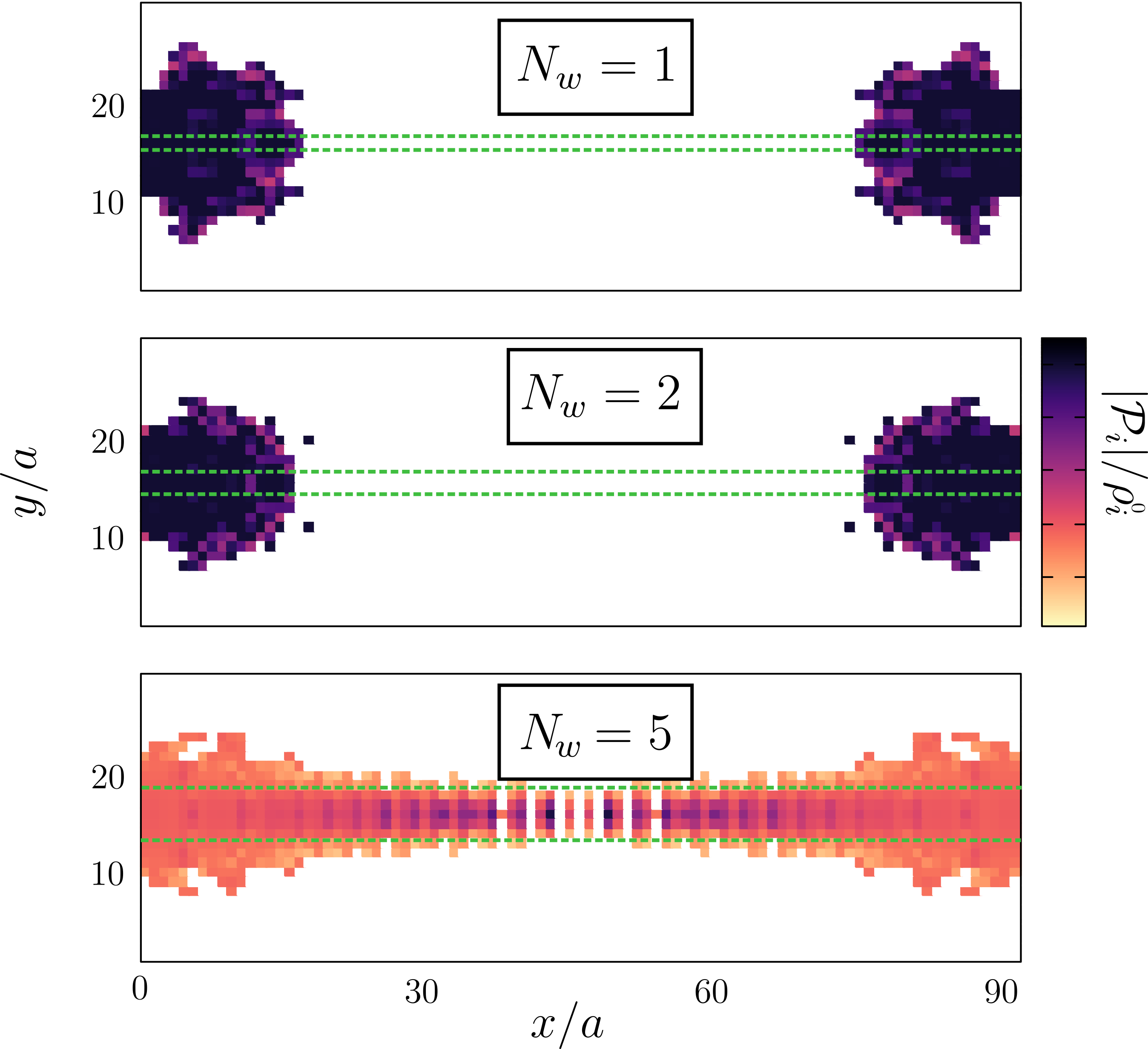}
\caption{The spatial profile of $|{\cal{P}}_{i}|/\rho^0_i$ for the potential zero-energy Majorana quasiparticles appearing in the metallic strip consisting of 1, 2 and 5 rows of atomic chains, as indicated. We have used the same model parameters as in Fig.~\ref{homogeneous}:  $\Delta=0.25t$, $\phi=\pi$, $\lambda=0.5t$, $B_{0}=0.1t$, $\mu=-3.75t$. The difference between the well isolated MBS and the delocalized states in the wider strip is evident.} 
\label{fig_rat}
\end{figure}

Although the spatial variation of the phase is not currently measurable we can compare the absolute value of $|{\cal{P}}_{i}|$ with $\rho_i(0)$. For a MBS these must be the same, therefore if we plot the ratio of these measurable quantities,  $|{\cal{P}}_{i}|/\rho_i(0)$, it should be flat for a MBS. In Fig.~\ref{fig_rat} we show results for $N_w\in\{1,2,5\}$, comparable to Fig.~\ref{homogeneous}, which show that the MBS profile is indeed flat. For $N_w=5$, when the two MBS at either end of the strip start to overlap, this quantity is no longer flat. Thus this can be used as an experimental determination of whether localized states are actually MBS. It is worth noting that increasing the length $N_x$ of the system would result in $|{\cal{P}}_{i}|/\rho_i(0)$ being flat even for $N_{w}>5$.

\begin{figure}[b!]
\includegraphics[width=0.95\linewidth,keepaspectratio]{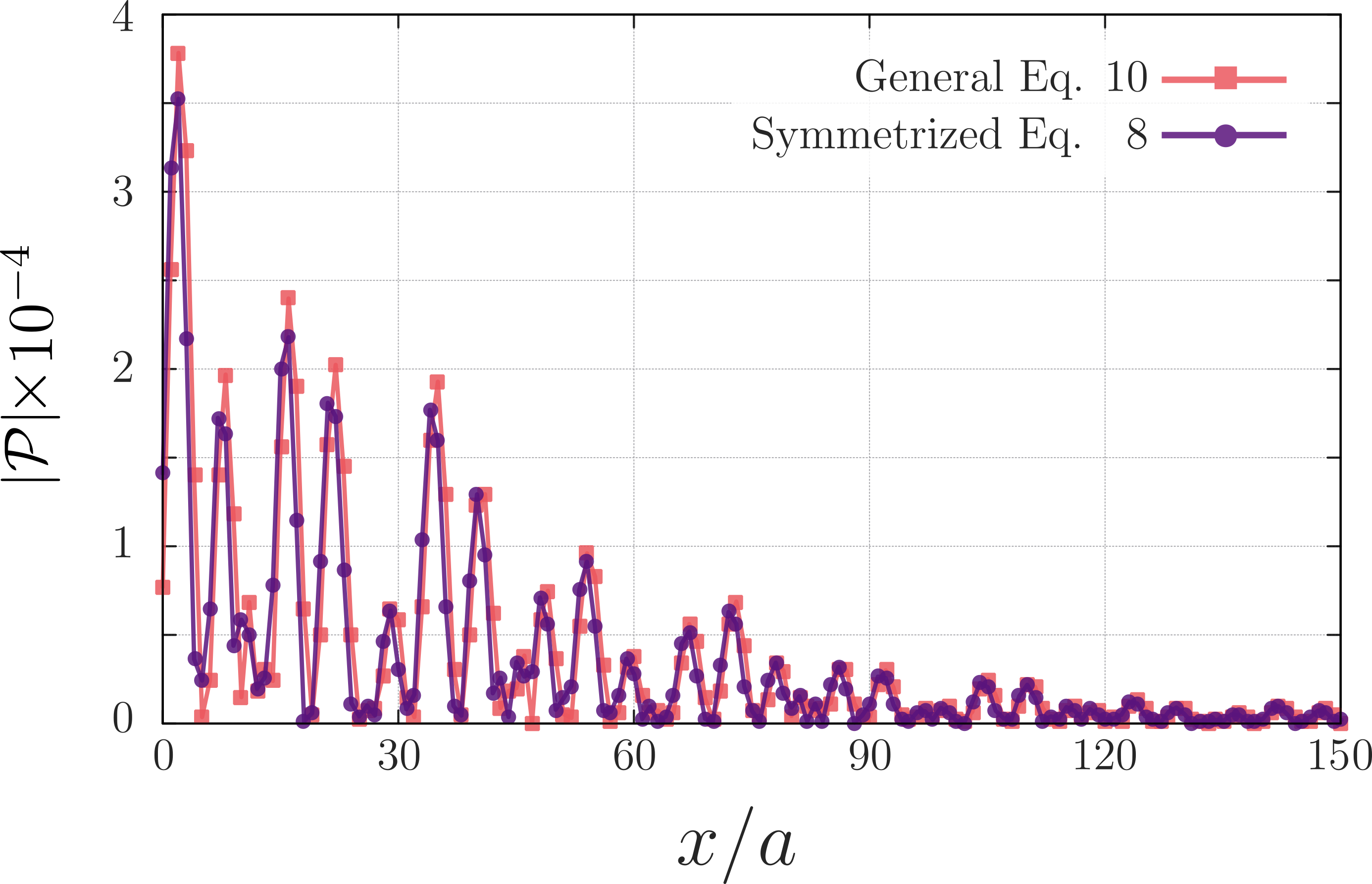}
\caption{Same as in Fig.~\ref{profiles} obtained by {\it Kwant} package for the larger size system $N_x=2000$, $N_y=300$ and $N_{w}=5$. In analogy to Refs.~\cite{Fornieri-2019,Ren-2019} we have assumed the model parameters $t=32$ meV, $\Delta=0.16$ meV, $\lambda=2.85$ meV, $B_0=0.12$ meV, $\mu=-120$ meV, imposing the phase difference $\phi=\pi$.} 
\label{profiles2}
\end{figure}

Our Eq.~(\ref{polarization_experiment}) demonstrates analytically the feasibility of measuring a modulus of the symmetrized Majorana polarization, which we have so far illustrated for relatively small system sizes (Fig.~\ref{profiles}). Now we would like to verify its usefulness for probing an experimentally realistic situation. For this purpose we have used the {\it Kwant} package \cite{Groth_2014} to compute the local (\ref{polarization_per_spin}) and symmetrized (\ref{averaged_polarization_per_spin}) Majorana polarizations for the larger sample, comprising $2000 \times 300$ sites, adopting the model parameters reported in Refs.~\cite{Fornieri-2019,Ren-2019}. Fig.~\ref{profiles2} displays profiles of these polarizations obtained near a boundary of the proximitized metallic strip. It clearly shows that in realistic samples the symmetrized Majorana polarization is practically identical with the local one.

\begin{figure*}
\includegraphics[width=0.95\linewidth,keepaspectratio]{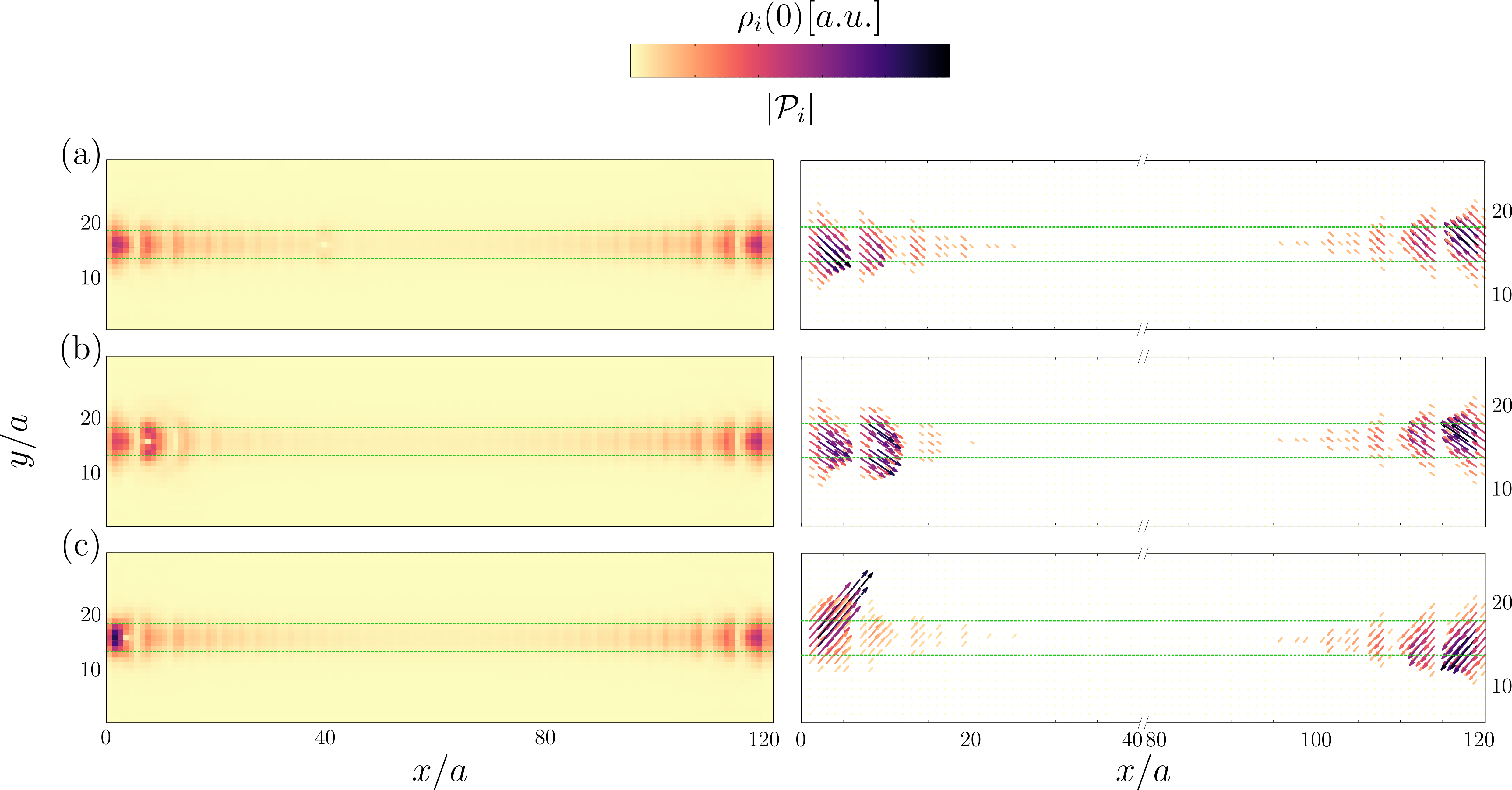}
\caption{The spatial density profile of the zero-energy Majorana quasiparticles (left panels) and their polarizations (right panels) obtained in the proximitized metallic strip, consisting of $N_{w}=5$ rows of atoms, in presence of a point-like electrostatic defect with $V_0=10t$ at site: (a) $i_0=40a$, (b) $i_0=8a$, and (c) $i_0=4a$, with $N_x=120$ and $N_y=30$.  The magnitude of the arrows in the right hand side plots show $|\mathcal{P}_{i\sigma}|$ and their direction show $\textrm{Arg}\,\mathcal{P}_{i\sigma}$. We note that the phase is only well defined up to a global shift.} 
\label{localization_center}
\end{figure*}

\section{Localization of the Majorana modes}
\label{sec.impurity}

In Sec.~\ref{sec.transition} we have shown that upon increasing the width $N_{w}$ of a metallic strip the topological superconducting state reveals (i) smearing of the Majorana quasiparticle, and (ii) development of the transverse gradient of the Majorana polarization. One may ask, however, whether there is any chance of localization of the Majorana modes. In this section we illustrate that such an effect could be observable due to local defects introduced in certain regions of the metallic strip.

Topological superconductivity in 1-dimensional wires and atomic chains has been shown to be robust against weak disorder~\cite{Brouwer_2011,DeGottardi_2013,Hui_2015,
Hoffman_2016,Zhang_2016,Hegde_2016,Cole_2016,Pekerten_2017}, noise~\cite{Hu_2015}, inhomogeneous spin-orbit coupling \cite{Klinovaja2015}, reorientation of the magnetic field~\cite{kiczek.ptok.17,Kaladzhyan2017a}, and thermal fluctuations~\cite{klinovaja.stano.13,Simon2015,Scalettar2015,maska_etal_2019}. Sufficiently strong scattering centers, however, could effectively break these 1-dimensional systems into separate segments, inducing additional pairs of the Majorana modes~\cite{maska.gorczyca.17}. Such a mechanism can be expected to be inefficient in 2-dimensional systems. To verify this conjecture for the quasi 2-dimensional heterostructure discussed in this paper we take into consideration a point-like electrostatic defect ${\cal{H}}_{imp}= V_{0} d_{i_{0}\sigma}^{\dagger} d_{i_{0}\sigma}$ positioned at site $i_{0}$ of the metallic region. We assume this scattering potential to be rather strong, $V_{0}=10t$, as otherwise its influence would be less visible.

Let us first assume the scattering potential to be placed in a central part of the metallic region (Fig.~\ref{localization_center}(a)). Neither
the zero-energy spectral function nor the Majorana polarization
reveal any influence of such an electrostatic impurity on the existing Majorana quasiparticles, in contrast to the properties of 1-dimensional topological superconductors~\cite{maska.gorczyca.17}. This behaviour seems to be quite natural, because the Majorana modes are safely distant from the impurity. 

Contrary to this situation, let us next consider the scattering potential near the left side of the metallic strip (Fig.~\ref{localization_center}(b,c)). We selected the specific sites $i_{0}=4$ and $i_{0}=8$, in order to guarantee a considerable overlap of the local defect with the left hand Majorana mode. Under such circumstances the scattering potential has a substantial influence both on the spectral function (left panels) and the Majorana polarization (right panels). This electrostatic impurity reduces the spatial extent of the Majorana quasiparticle on the left hand side, whereas the other Majorana quasiparticle is practically left intact. Such a tendency towards localization of the Majorana modes has been recently predicted by Haim and Stern~\cite{Stern-2019_disorder}, when investigating the different role of weak extended disorder. 

Our present study provides more detailed information concerning such disorder-induced-localization, indicating that: (i) disorder present in the internal segments of the metallic strip would naturally be rather ineffective for the Majorana quasiparticles, whereas (ii) disorder introduced to the regions of already existing Majorana quasiparticles substantially reduces their spatial extent. The considerations discussed in Sec.~\ref{SESAR} suggest that empirical detection of this subtle phenomenon could be feasible. A tendency towards the localization of the Majorana quasiparticles could be observed in the maps of differential conductance for the spin polarized Andreev current induced via the metallic region in the presence of the intentionally deposited local defects. Such an electrostatic scattering potential could be created by applying gate potentials, whereas a magnetic potential, leading to similar effects, can be obtained by locally perturbing the Zeeman field.

\section{Summary}
\label{sec.sum}

We have theoretically studied the properties of the Majorana quasiparticles emerging in a narrow metallic strip sandwiched between two $s$-wave superconductors in a Josephson-junction geometry. The topological superconducting phase has been recently reported for such metallic strips by the groups in Copenhagen \cite{Fornieri-2019} and Harvard \cite{Ren-2019} with a length-to-width ratio ranging from $20$ to $100$, respectively. Using the Bogoliubov de-Gennes treatment we have investigated the role of the finite metallic strip width, exploring its influence on: (a) spatial profiles of the zero-energy quasiparticles and (b) topography of the Majorana polarization that probes the particle-hole overlap of the zero-energy quasiparticles. Furthermore, we have proposed a feasible method for detecting the magnitude of such a Majorana polarization by measuring the differential conductance in spin-polarized Andreev reflection spectroscopy.

We have also analyzed the influence of strong (point-like) electrostatic defects on the Majorana modes. We have revealed that such a local scattering potential can affect the localization length of the Majorana quasiparticles if deposited near the ends of the metallic strip. Under such circumstances the neighboring Majorana mode substantially reduces its spatial extent, which can be compared with what has been predicted in Ref.~\cite{Stern-2019_disorder}, whereas the opposite-end Majorana mode remains practically intact. Similar coexistences of the localized and delocalized Majorana quasiparticles have been previously observed by scanning tunneling spectroscopy using a disordered monolayer of superconducting $Pb$ coupled to underlying $Co-Si$ magnetic islands~\cite{PascalSimon_etal.18}. We hope that proximitized metallic strips would be a convenient platform not only for the realization of topological superconductivity, tunable by the magnetic field and Josephson phase, but could also allow for manipulating the Majorana quasiparticle length-scale. 

\begin{acknowledgments}
We thank Benedikt Scharf for instructive discussions. This work is supported by National Science Centre (Poland) under the grants UMO-2017/27/N/ST3/01762 (SG), UMO-2018/29/B/ST3/01892 (NS), 
and UMO-2017/27/B/ST3/01911 (TD). 
\end{acknowledgments}

\appendix

\section{Influence of the Josepshon phase}
\label{phasal_effect}

In the main part of this manuscript we have analyzed the spectroscopic properties of the Majorana quasiparticles, focusing on the particular case $\phi=\pi$, that is optimal for occurrence of the topological superconducting state. Similar qualitative features would be also observable for other values of the Josephson phase $\phi\neq\pi$, provided that model parameters (such as, e.g., the magnetic field) are appropriately tuned \cite{Pientka-2017,Hell-2017}. A possible criterion (proposed by one of us \cite{Sedlmayr2015b,Sedlmayr2016}) for determination of the topological diagram relies on the Majorana polarization for MBS to be non-vanishing when summed over a portion of the investigated system where the bound states should reside, in fact it should be equal to the total density of the state in the same region. Therefore
\begin{equation}
    \mathcal{P}=\sum_{i\in\mathcal{R}}\frac{|\mathcal{P}_{in_0}|}{\rho_i(E_{n_0})}=1
\end{equation}
for a MBS. This serves as a proxy for being in the topologically non-trivial phase. In the present scenario one can choose for this purpose either the leftmost or rightmost quarter of the metallic strip as the region $\mathcal{R}$. More details can be found in Ref.~\onlinecite{Sedlmayr2016}.

\begin{figure}
\includegraphics[width=0.95\linewidth,keepaspectratio]{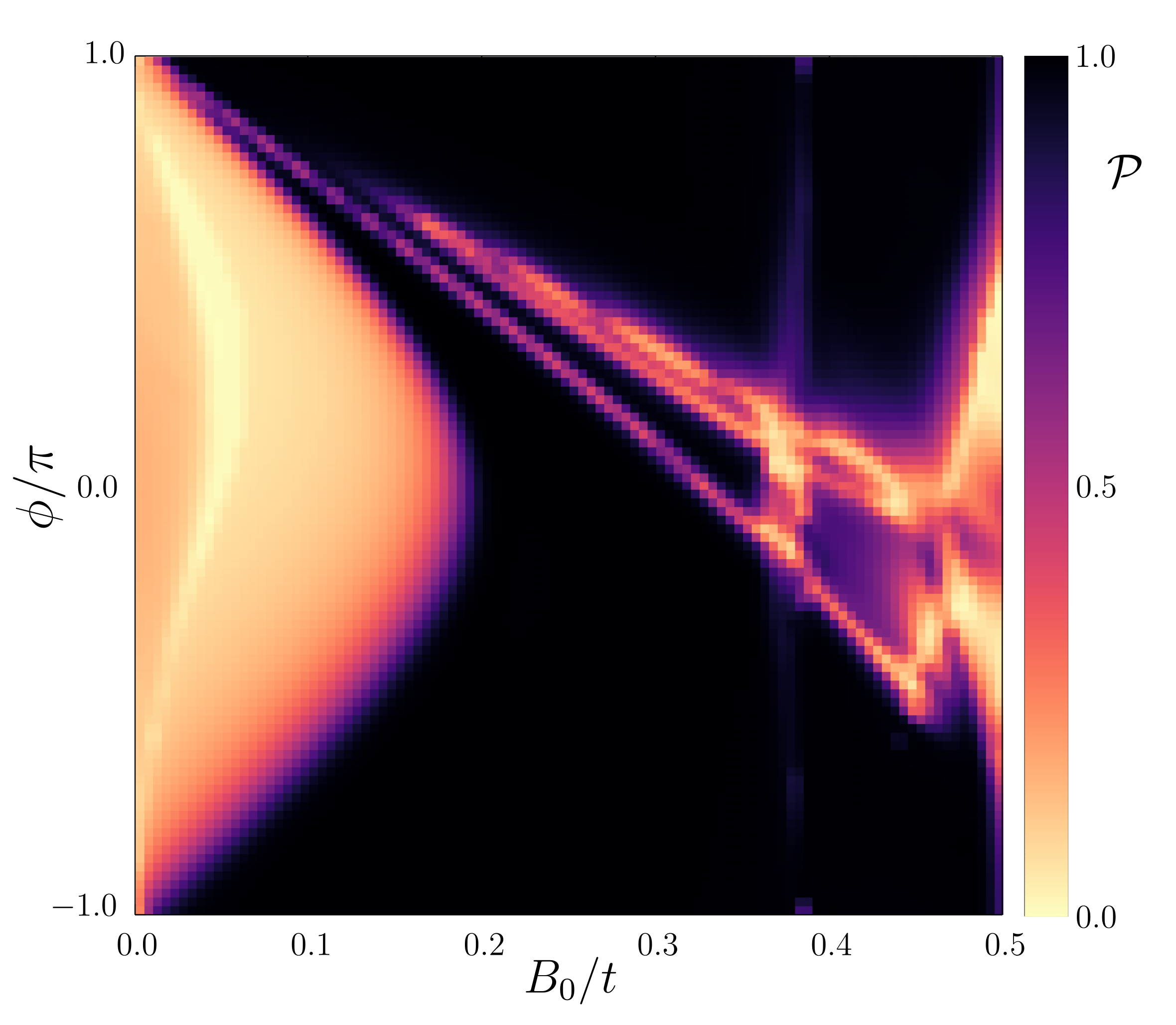}
\caption{The topological superconducting phase (dark area) with respect to the magnetic field $B_{0}$ and the Josephson phase $\phi$  obtained by the criterion concerning the global Majorana polarization~\cite{Sedlmayr2015b,Sedlmayr2016}. Numerical computations were done for $N_{w}=2$ and the same model parameters as in Fig.~\ref{homogeneous}.}
\label{phasediag}
\end{figure}

Fig.~\ref{phasediag} displays the topological phase diagram, with the dark area being the nontrivial phase, obtained using this criterion for our heterostructure. We show the phase diagram with respect to the magnetic field $B_{0}$ and the Josephson phase $\phi$. Let us remark, that such a criterion yields a smooth changeover between the topologically trivial and nontrivial superconducting states instead of a sharp transition. It can be hence useful for exploring the robustness of the topological state against perturbations such as inhomogeneity or thermal fluctuations.

\section{Localization by disorder}
\label{disorder_effect}

\begin{figure*}
\includegraphics[width=0.95\linewidth,keepaspectratio]{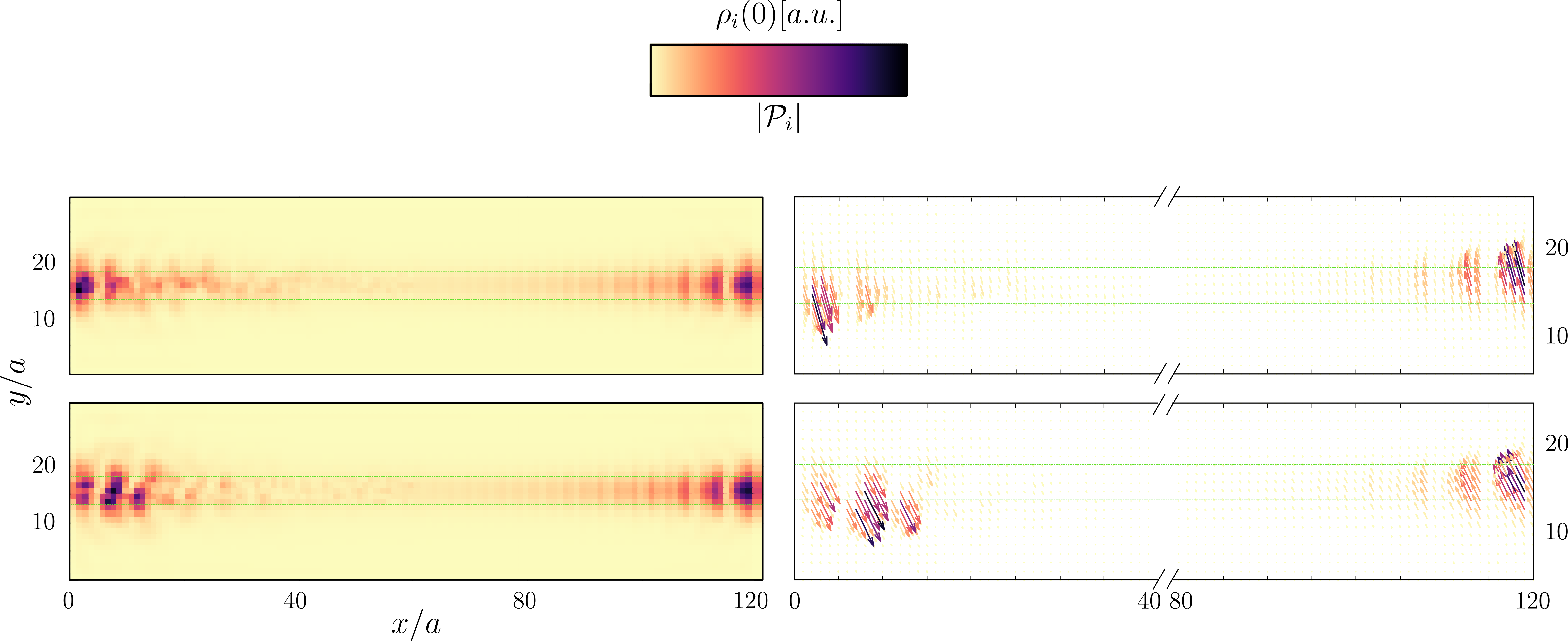}
\caption{The zero-energy spectral function $\rho_{i}(0)$ and Majorana polarization with the random scattering potential in the left half of the proximitized strip $V_{i} \in \left( -\delta V ; \delta V \right)$ with $\delta V=0.5t$ (top) and $\delta V=0.75t$ (bottom). Results are obtained for the same model parameters as in Fig.~\ref{homogeneous}.}
\label{disorder}
\end{figure*}

Numerous studies ~\cite{Brouwer_2011,DeGottardi_2013,Hui_2015,Hoffman_2016,Zhang_2016,Hegde_2016,Cole_2016,Pekerten_2017} have shown, that the topological superconductivity of one-dimensional systems is pretty stable against a weak disorder. In the phase controlled Josephson junctions, however, Haim and Stern~\cite{Stern-2019_disorder}
have predicted that a weak disorder could affect topography of the Majorana modes by reducing their spatial extent. We address here briefly this intriguing effect within the Bogoliubov de-Gennes formalism. For the purpose of comparison with the results discussed in Sec.~\ref{sec.impurity} we assume that the random scattering potential $V_{i}$ is present only in the left hand side of the metallic strip, whereas the remaining (right hand side) part is homogeneous.

Fig.~\ref{disorder} displays topographic features of the zero-energy quasiparticles obtained for representative amplitudes $\delta V$ of the random scattering potential $V_{i} \in \left( -\delta V ; \delta V \right)$. The upper panel corresponds to $\delta V = 0.5t$ and the bottom one to $\delta V = 0.75 t$, respectively. We can clearly recognize shrinking of the Majorana quasiparticle extent in a disordered fragment of the metallic strip while in the other (homogeneous) piece the Majorana quasiparticle remains practically untouched. Such localization driven by the strong local scattering potential  (Sec.~\ref{sec.impurity}) and the weak disorder \cite{Stern-2019_disorder} seems to be unique to the topological superconducting phase in quasi two-dimensional systems.

\bibliography{biblio.bib}

\end{document}